# RURANET++: An Unsupervised Learning Method for Diabetic Macular Edema Based on SCSE Attention Mechanisms and Dynamic Multi-Projection Head Clustering

Wei Yang[1], Yiran Zhu[1], Jiayu Shen[2], Yuhan Tang[2], Chengchang Pan[2], Hui He[3], Yan Su[2,*], Honggang Qi[2,*]

[1] Department of computer, North China Electric Power University (Baoding)
[2] School of Computer Science and Technology, University of Chinese Academy of Sciences
[3] Faculty of Art and Science; Beijing Normal University
1227698971@qq.com
hgqi@ucas.ac.cn

**Abstract.** Diabetic Macular Edema (DME), a prevalent complication among diabetic patients, constitutes a major cause of visual impairment and blindness. Although deep learning has achieved remarkable progress in medical image analysis, traditional DME diagnosis still relies on extensive annotated data and subjective ophthalmologist assessments, limiting practical applications. To address this, we present RURANET++, an unsupervised learning-based automated DME diagnostic system. This framework incorporates an optimized U-Net architecture with embedded Spatial and Channel Squeeze & Excitation (SCSE) attention mechanisms to enhance lesion feature extraction. During feature processing, a pre-trained GoogLeNet model extracts deep features from retinal images, followed by PCA-based dimensionality reduction to 50 dimensions for computational efficiency. Notably, we introduce a novel clustering algorithm employing multi-projection heads to explicitly control cluster diversity while dynamically adjusting similarity thresholds, thereby optimizing intra-class consistency and inter-class discrimination. Experimental results demonstrate superior performance across multiple metrics, achieving maximum accuracy (0.8411), precision (0.8593), recall (0.8411), and F1-score (0.8390), with exceptional clustering quality. This work provides an efficient unsupervised solution for DME diagnosis with significant clinical implications.

**Keywords:** Diabetic Macular Edema; Deep Learning; Unsupervised Learning

## 1 Introduction

The World Health Organization's 2024 report reveals a dramatic surge in global diabetes prevalence from 7% to 14% among adults between 1990-2022, affecting over 800 million individuals with approximately 59% lacking effective treatment [1-2]. As a common diabetic complication, Diabetic Macular Edema (DME) represents a leading cause of irreversible vision loss [3]. Early detection and intervention could substantially mitigate visual deterioration, yet conventional diagnosis remains constrained by subjective ophthalmologist evaluations and operational inefficiencies.

While deep learning has advanced medical image analysis [4-5], current methodologies face dual challenges: Supervised approaches (e.g., lesion segmentation [6-8], self-supervised diagnosis [9-10]) require labor-intensive annotated data constrained by medical expertise and costs. Simultaneously, existing methods suffer from insufficient discriminative power and noise sensitivity during feature extraction and clustering, limiting lesion differentiation accuracy.

To address these limitations, we propose an unsupervised diagnostic framework for diabetic macular edema (DME) that simultaneously optimizes image segmentation and feature clustering through joint learning. Departing from conventional approaches, our method introduces a dynamic multi-projection clustering mechanism with adaptive similarity thresholding to maintain cluster diversity.

Our principal contributions are threefold: (1) We develop an unsupervised DME diagnostic system that reduces annotation dependency and improves diagnostic accuracy; (2) We integrate SCSE attention mechanisms into the U-Net architecture and identify an optimal PCA dimensionality (50 dimensions) that balances computational efficiency with critical feature retention; (3) Implementation of multi-projection head clustering that ensures cluster stability and accuracy through diversity control and dynamic similarity threshold adaptation.

## 2 Methodology

### 2.1 Hierarchical Lesion Segmentation Network

We propose a multi-scale segmentation model based on an enhanced U-Net architecture to improve encoder performance. Candidate models are constructed by integrating diverse backbones into the U-Net decoder, with optimal backbones selected per lesion type through comparative experiments. An SCSE module [12] is embedded to enhance feature discriminability for subtle lesions (detailed in Section 3.2 , Fig. 1a).

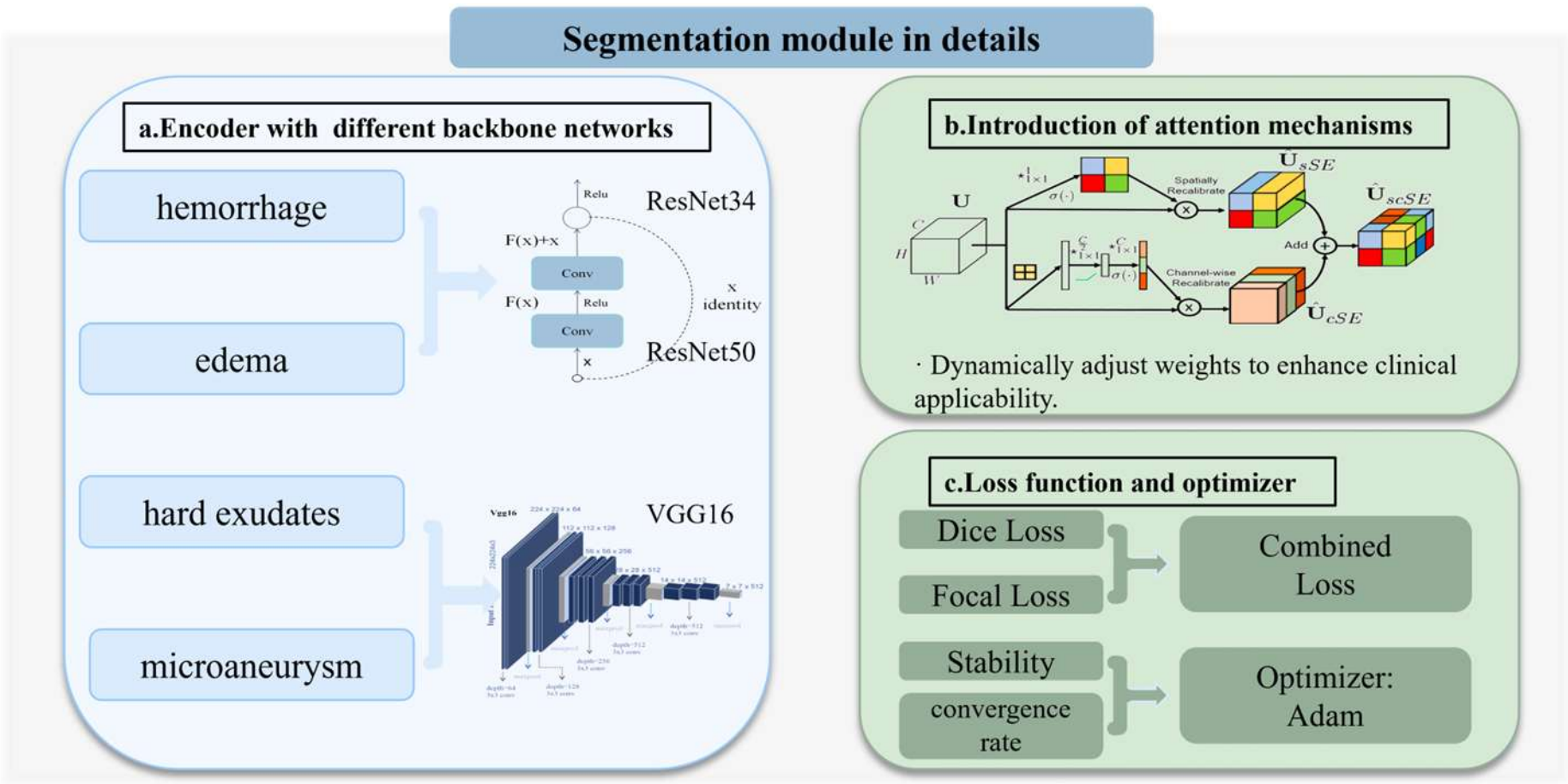

**Fig. 1.** The process of designing our focal segmentation module

**Lesion-Specific Backbone Network Selection.** To address DME lesion heterogeneity, we propose an adaptive backbone selection strategy for U-Net encoder, selecting optimal backbones independently for each of four target lesions. Ten mainstream backbones (e.g., VGG19) were evaluated, with optimal backbones determined by individual lesion segmentation performance (IoU, accuracy, precision) (detailed in Section 3.2).

**SCSE Attention Mechanism Integration in U-Net.** The SCSE module enhances feature representation by jointly optimizing spatial and channel-wise feature responses (Fig. 1b). Embedded in U-Net skip connections, it processes encoder features $F_e \in R^{H\times W\times C}$ and decoder features $F_d \in R^{H\times W\times C}$ as follows: 1) Concatenates $F_e$ and $F_d$ to form $F_{cat} \in R^{H\times W\times 2C}$; 2) Computes spatial attention $S \in R^{H\times W\times 1}$ and channel attention $C \in R^{1\times 1\times C}$; 3) Recalibrates features via S and C to output $F_{out} \in R^{H\times W\times C}$.

**Loss Function Design.** To address class imbalance (dominant background pixels vs. sparse lesion pixels), a hybrid CombinedLoss (Fig. 1c) integrating Dice Loss and Focal Loss (1) is proposed. This dual-objective function balances global segmentation accuracy and sensitivity to small lesions:

$$FocalLoss = -\alpha_t(1 - p_t)^{\gamma}\log(p_t) \quad (1)$$

Training employs the Adam optimizer (learning rate = $1e^{-4}$, $\beta_1$ =0.9, $\beta_2$ =0.999) with early stopping based on validation Dice coefficient.

## 2.2 Multi-projection Clustering Framework

Following lesion segmentation, we employ a pre-trained GoogLeNet model to extract features from retinal images and implement a dynamic multi-projection head clustering algorithm for disease image classification (Fig. 2).

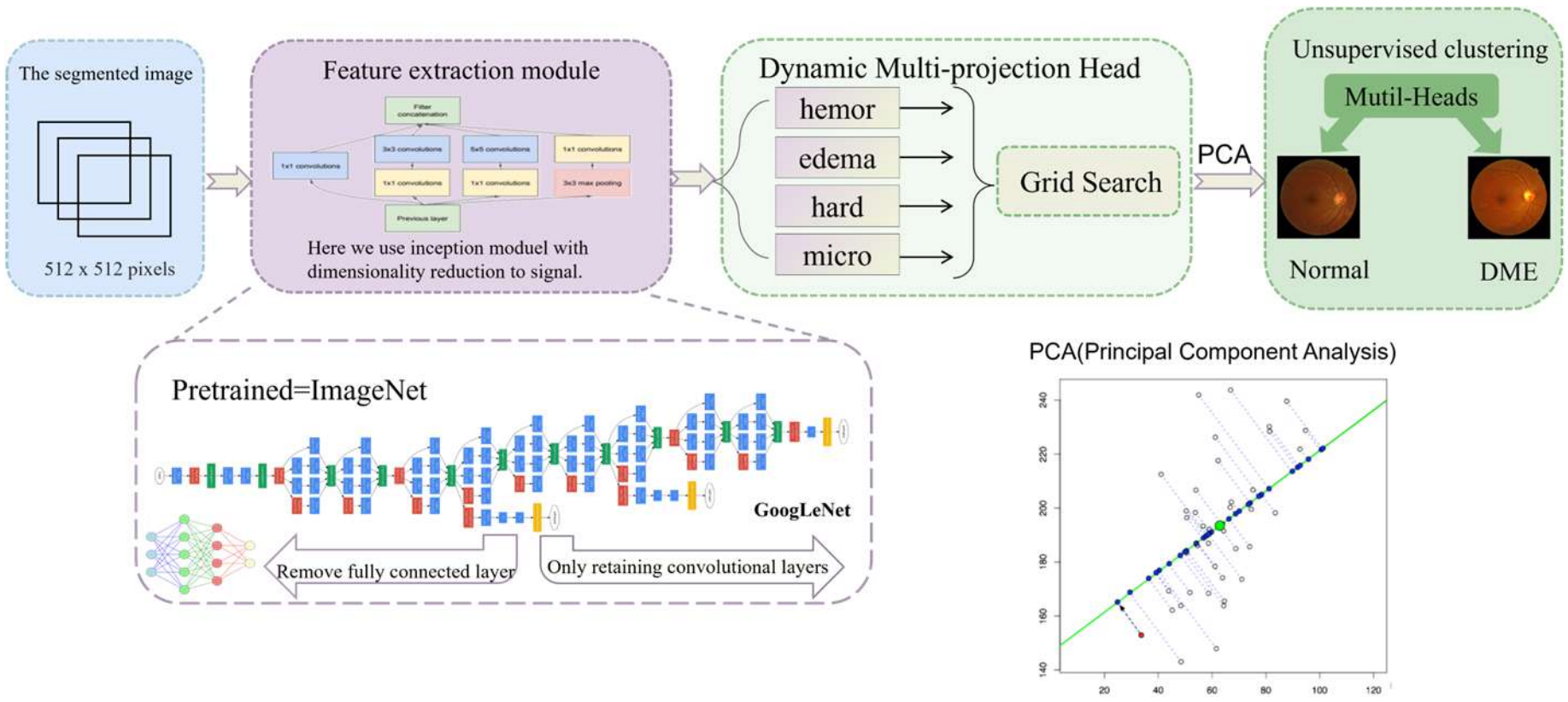

**Fig. 2.** Architecture of the proposed clustering module

**Multi-lesion Feature Weighting Strategy.** To quantify lesion-specific contributions to diabetic retinopathy grading, learnable weights $w_i \in [1.0,10.0]$ are assigned to four lesion features (hard exudates, microaneurysms, disk hemorrhages, edema). Optimal weights are determined via grid search over $\{1.0,5.0,10.0\}^4$, maximizing clustering accuracy (2) . Aggregated features are computed as:

$$F_{global} = \frac{1}{4}\sum_{i=1}^{4} w_i \cdot f_i \ (i = 1,2,3,4) \tag{2}$$

**Deep Feature Extraction.** We utilize an ImageNet-pretrained GoogLeNet (Inception-v1) as the feature extractor, removing the original classification layer while retaining the feature extraction backbone. This leverages ImageNet-pretrained weights for generalization and captures semantic features for simplified clustering.

**Dynamic Multi-projection Head Clustering.** Building upon the multi-lesion feature weighting strategy, we design a dynamic multi-projection head clustering algorithm to enhance the diversity of clustering results across distinct lesions and guide the model to focus on lesion-specific pathological characteristics. Specifically, after unified feature extraction through the backbone network of the clustering model, lesion-specific features are processed by dedicated projection heads to generate discriminative representations tailored to each lesion type. These refined features undergo weighted aggregation followed by K-Medoids clustering. Recognizing the varying clinical significance of different lesions in diabetic macular edema (DME) diagnosis[19-20], we employ a weight exploration network to evaluate the impact of different weighting schemes on clustering performance, thereby quantifying the model's attention allocation to specific lesions.

During training, a dual-term objective function optimizes clustering quality and lesion-specific diversity:

$$\mathcal{L}_{total} = \mathcal{L}_{div} + \lambda \mathcal{L}_{\boldsymbol{cluster}} \tag{3}$$

where $\lambda = 0.3$ denotes the balancing coefficient. The diversity loss term incorporates a ReLU-activated cosine similarity penalty:

$$\mathcal{L}_{div} = \frac{1}{C_4^2}\sum_{i<j} \text{ReLU}\left(\frac{1}{2N}\sum_{n=1}^{N}\left(max\left(S_{ij}^{(n)}\right) + max\left(S_{ji}^{(n)}\right)\right) - T\right) \quad (4)$$

The clustering loss term is derived from the K-Medoids algorithm:

$$\mathcal{L}_{cluster} = \frac{1}{4}\sum_{i=1}^{4} E\left[\min\left(\left\|f_i - m_1^{(i)}\right\|_2, \left\|\boldsymbol{f}_i - \boldsymbol{m}_2^{(i)}\right\|_2\right)\right] \quad (5)$$

where $m_k^{(i)}$ represents the k-th cluster centroid for the i-th lesion feature.
**Multidimensional Metric-based Feature Dimension Reduction Optimization.** To balance feature expressiveness and computational efficiency, we propose a PCA-based reduction strategy optimized via multidimensional metric evaluation:
*Cumulative variance contribution ratio.* Evaluates global information preservation by the first k principal components. Experimental results show a cumulative ratio of 94.32% at $k = 50$.
*Reconstruction error quantification.* Measures information loss through projection-inverse projection :

$$ReconstructionError = \frac{1}{N}\sum_{i=1}^{N}\left\|X^{(i)} - W_k W_k^T X^{(i)}\right\|_2^2 \quad (6)$$

where $W_k \in R^{d\times k}$ represents the projection matrix. At $k = 50$, the reconstruction error reaches 0.0568, balancing computational efficiency and information preservation.

### 2.3 Evaluation Protocol

To comprehensively assess the model's performance in fundus lesion segmentation and clustering tasks, we establish a multidimensional evaluation framework spanning supervised segmentation and unsupervised clustering metrics.
**Segmentation Metrics.** Accuracy, Precision, Recall, and IoU.
**Clustering Metrics.** A dual-metric system combining the Calinski-Harabasz (CH) Index and Davis-Bouldin (DB) Index is implemented.

The orthogonal dual-metric framework mitigates biases and robustly evaluates the applicability of unsupervised clustering in fundus image analysis.

## 3 Experiments

### 3.1 Data and Preprocessing

This study employs a multi-source fundus image dataset, derived from six publicly available resources: Retinal Lesions[13]、Retinal Vessel Segmentation[14]、Retinal

Vessel Segmentation Combined[15]、REFUGE2[16]、IDRI[17]、Final Clean Haemorrhage[18] datasets. The integrated dataset comprises fundus images from 302 patients with diabetic macular edema (DME), maintaining a balanced distribution of 151 positive and 151 negative samples. Each image is annotated with pixel-level masks for four critical pathological lesions: hard exudates, microaneurysms, optic disc hemorrhages, and edema regions.

Preprocessing Pipeline. All images were resampled to a uniform resolution of 512×512 pixels using bicubic interpolation, followed by pixel value normalization to the range [0, 1]. The dataset was partitioned into training, validation, and test sets at an 8:1:1 ratio, with patient ID to ensure data independence across subsets.

## 3.2 Comparison of Backbone Networks in the Segmentation Module

The section compares different backbone networks to selecte optimal backbone networks for each lesion type to improve segmentation accuracy (Table 1). Through comprehensive comparisons, these are final feature extraction networks (Table 2).

**Table 1.** Comparison of backbone networks for different lesions

| Model | Pretrained | Metrics | | | | | |
|---|---|---|---|---|---|---|---|
| | | Microaneurysm | | | Hard exudates | | |
| | | IoU | Accuracy | Precision | IoU | Accuracy | Precision |
| **ResNet18** | ImageNet | 0.0173 | 0.9902 | 0.1367 | 0.8548 | **0.9971** | 0.9528 |
| **ResNet34** | | 0.0663 | 0.9827 | 0.1112 | 0.8631 | **0.9973** | **0.9658** |
| **ResNet50** | | 0.0479 | 0.9826 | 0.0920 | 0.8623 | 0.9965 | **0.9541** |
| **VGG16** | | **0.2566** | **0.9910** | **0.4498** | **0.8719** | **0.9968** | **0.9625** |
| **VGG19** | | **0.2264** | **0.9921** | **0.5342** | 0.8658 | **0.9969** | **0.9651** |
| **MobileNet** | | 0.0302 | 0.9809 | 0.0651 | 0.8041 | 0.9958 | 0.9111 |
| **DenseNet121** | | 0.0475 | 0.9826 | 0.1065 | 0.8372 | 0.9966 | 0.9419 |
| **DenseNet161** | | 0.0543 | 0.9846 | 0.1099 | 0.8633 | **0.9971** | 0.9502 |
| **DenseNet169** | | 0.0583 | 0.9854 | 0.1282 | 0.8664 | **0.9975** | **0.9634** |
| **DenseNet201** | | 0.0537 | 0.9820 | 0.0973 | 0.8714 | **0.9975** | 0.9576 |

**Table 2.** Optimal modules for segmentation tasks

| Foci | Pretrained | Backbone network | Metrics | | |
|---|---|---|---|---|---|
| | | | IoU | Accuracy | Precision |
| **Hemorrhage** | ImageNet | ResNet34 | 0.4942 | 0.9956 | 0.8008 |
| **Edema** | | ResNet50 | 0.9925 | 0.9867 | 0.9988 |
| **Hard exudation** | | VGG16 | 0.8719 | 0.9968 | 0.9625 |
| **Microaneurysm** | | VGG16 | 0.2566 | 0.9910 | 0.4498 |

## 3.3 Clustering Algorithms

**Supervised Binary Classification with Different Backbone Networks.** This section demonstrates the differences in lesion recognition when using different backbone networks. We select optimal models by monitoring validation performance (Table 3).

**Comparison of Different Backbone Networks and Clustering Methods under Supervised Learning.** Based on Section 3.3.1 results, backbone networks significantly impact model performance. Therefore, this section compares eight backbone networks and four clustering methods—as detailed in Table 4, with DMH + GoogLeNet demonstrated excellent performance in accuracy (0.8411), precision (0.8593), recall (0.8411), F1-score (0.8390) and high clustering quality, so it is the final model.

**Table 3.** Impact of different backbone networks

| Backbone network | Metrics | | | | |
|---|---|---|---|---|---|
| | Accuracy | Precision | Recall | F1-Score | ROU AUC |
| AlexNet | 0.6452 | 0.5918 | 0.9355 | 0.7250 | 0.8085 |
| GoogLeNet | 0.7258 | 0.7333 | 0.7097 | 0.7213 | 0.8044 |
| MobileNet | 0.7903 | 0.8214 | 0.7419 | 0.7797 | 0.8554 |
| ResNet18 | 0.7581 | 0.9000 | 0.5806 | 0.7059 | 0.8200 |
| ResNet34 | 0.6290 | 1.0000 | 0.2581 | 0.4103 | 0.7700 |
| ResNet50 | 0.5484 | 1.0000 | 0.0968 | 0.1765 | 0.6046 |

**Table 4.** Different Backbone Networks and Clustering Methods under Supervised Learning

| Method | Backbone networks | Metrics | | | | | |
|---|---|---|---|---|---|---|---|
| | | Accuracy | Precision | Recall | F1-Score | Calinski-Harabasz | Davies-Bouldin |
| k-medoids | AlexNet | 0.8333 | 0.8437 | 0.8333 | 0.8321 | 15.9312 | 2.2379 |
| | GoogLeNet | 0.8600 | 0.8606 | 0.8600 | 0.8599 | 36.3945 | 1.8753 |
| | ResNet18 | 0.8333 | 0.8437 | 0.8333 | 0.8321 | 161.6702 | 0.7050 |
| | VGG16 | 0.8133 | 0.8283 | 0.8133 | 0.8112 | 12.4736 | 3.1007 |
| Kmeans | GoogLeNet | 0.8000 | 0.8054 | 0.8000 | 0.7991 | 267.8952 | 0.5985 |
| | ResNet34 | 0.8333 | 0.8437 | 0.8333 | 0.8321 | 19.4077 | 2.6955 |
| | VGG16 | 0.7667 | 0.7749 | 0.7667 | 0.7649 | 12.0028 | 3.1943 |
| | VGG19 | 0.8067 | 0.8134 | 0.8067 | 0.8056 | 14.0618 | 2.7996 |
| Agg | AlexNet | 0.8267 | 0.5630 | 0.5511 | 0.5539 | 14.7440 | 2.1688 |
| | GoogLeNet | 0.8400 | 0.5742 | 0.5600 | 0.5626 | 16.1686 | 2.1582 |
| | ResNet18 | 0.7267 | 0.5265 | 0.4844 | 0.4982 | 71.7643 | 1.5988 |
| | VGG16 | 0.8533 | 0.5780 | 0.5689 | 0.5728 | 11.0147 | 2.2108 |
| DMH | GoogLeNet | 0.8411 | 0.8593 | 0.8411 | 0.8390 | 142.8781 | 1.3839 |
| | AlexNet | 0.7517 | 0.8133 | 0.7517 | 0.7388 | 19.3640 | 3.4652 |
| | ResNet50 | 0.7583 | 0.7915 | 0.7583 | 0.7512 | 85.7031 | 1.5450 |
| | MobileNet | 0.5993 | 0.6172 | 0.5993 | 0.5834 | 299.5588 | 0.8191 |
| | VGG16 | 0.8245 | 0.8477 | 0.8245 | 0.8385 | 17.5721 | 4.0038 |
| | VGG19 | 0.7483 | 0.8027 | 0.7483 | 0.7365 | 33.6725 | 2.7154 |
| | ResNet18 | 0.8113 | 0.8289 | 0.8113 | 0.8087 | 294.3219 | 0.8433 |
| | ResNet34 | 0.8046 | 0.8241 | 0.8046 | 0.8169 | 73.3143 | 1.5340 |

### 3.4 Ablation Studies

Our work incorporates three major contributions: a pre-clustering segmentation mechanism, an embedded SCSE module in the segmentation network, and a dynamic multi-projection head clustering algorithm (DMH) . Ablation studies analyze each component's impact, starting from K-Medoids baseline, incrementally adding components, and finally removing PCA. Results are in Table 5.

**Baseline Model (K-Medoids).** As shown in Table 5a, the original K-Medoids algorithm without feature engineering achieves low accuracy (0.5430) and F1-score (0.5767), with high recall (0.6225) but low precision (0.5371), indicating many false positives. The Davies-Bouldin index (0.3350) also shows poor separability.

**Segmentation Mechanism.** As shown in Table 5b, introducing a GoogLeNet-based segmentation network (without SCSE) improves accuracy to 0.8400 and F1-score to 0.8397. The DB improves to 2.0853, confirming enhanced intra-class consistency.

**SCSE Mechanism in the Segmentation Network.** As shown in Table 5c, integrating the SCSE module (Table 5c) leads to qualitative improvements: accuracy surpasses the 0.86 threshold (+2.38%), and the F1-score increases by 2.41%.

**Dynamic Multi-Projection Head Clustering Algorithm.** As shown in Table 5d, the full experiment incorporates our novel DMH algorithm. Results show excellent performance across all metrics: accuracy (0.8411), precision (0.8593), recall (0.8411), and F1-score (0.8390). In clustering quality, the CH (142.8781) increases by 2.93-fold, and the DB (1.3839) decreases by 26.2%, indicating enhanced structural separability in the feature space, effectively capturing the heterogeneity of diabetic macular edema.
**Removal of PCA Dimensionality Reduction from the Full Model.** To evaluate the balance between computational efficiency and performance, we compare the full model with a PCA-free version. As shown in Table 5e, quantitative analysis reveals that PCA retains critical classification information while optimizing computation: accuracy (0.8377 vs. 0.8411) and F1-score (0.8355 vs. 0.8390) differ by less than 0.5%.

**Table 5.** Ablation studies of different variants

| Types | Metrics | | | | | |
|---|---|---|---|---|---|---|
| | Accuracy | Precision | Recall | F1_Score | calinski_harabasz_score | davies_bouldin_score |
| a | 0.5430 | 0.5371 | 0.6225 | 0.5767 | 2359.7505 | 0.3350 |
| b | 0.8400 | 0.8422 | 0.8400 | 0.8397 | 29.7464 | 2.0853 |
| c | 0.8600 | 0.8606 | 0.8600 | 0.8599 | 36.3945 | 1.8753 |
| d | 0.8411 | 0.8593 | 0.8411 | 0.8390 | 142.8781 | 1.3839 |
| e | 0.8377 | 0.8569 | 0.8377 | 0.8355 | 39.0020 | 2.6048 |

## 4 Discussion and Conclusion

The unsupervised learning framework, RURANET++, proposed in this study, significantly reduces the reliance on annotated data for the diagnosis of Diabetic Macular Edema (DME) by integrating an optimized U-Net segmentation network, the SCSE attention mechanism, and a dynamic multi-projection head clustering algorithm. The SCSE module effectively enhances the feature capture capability for minute lesions, while PCA dimensionality reduction optimizes computational efficiency by retaining 94.32% of the feature information. Additionally, the dynamic clustering mechanism improves the stability and accuracy of results through adaptive threshold adjustments.

The results show that the model's metrics in both segmentation and clustering tasks significantly outperform existing methods, validating the potential of unsupervised strategies in medical image analysis. This framework provides an efficient solution for early screening of DME in resource-limited settings. Future work will extend this research to include multi-modal data fusion and cross-disease generalization studies.

However, despite achieving favorable results across multiple performance metrics, the study still encounters certain computational complexities when handling high-dimensional data. To further enhance computational efficiency and model scalability, future research could explore more efficient feature dimensionality reduction techniques and optimize clustering algorithms to reduce computational overhead. Moreover, as the dataset size increases, the model's generalization ability will require further validation.